\documentclass[final]{draft}\usepackage{gnrl,grkl,math,phys}

\begin{document}

\begin{cover}
	\cID{~}
	\cTitle{Controlling the Flavour Changing Neutral Couplings of Multi-Higgs Doublets Models through Unitary Matrices}
	\cAuthor{a}{Jo\~ao M. Alves}{Joao.Aparicio@uv.es}
	\begin{cDeps}
		\cUni{a}{Departament de F\'isica Te\`orica and Instituto de F\'isica Corpuscular (IFIC)}
						{Universitat de Val\`encia -- CSIC, E-46100 Valencia, Spain}
	\end{cDeps}
	\begin{abstract}
		In this paper, we introduce Unitary Flavour Violation to produce Multi-Higgs Doublets Models where all flavour par%
		ameters are contained within three unitary matrices.
		After that, we identify two of its subclasses, the left and right models, which have naturally suppressed tree-lev%
		el Flavour Changing Neutral Couplings that easily avoid the experimental constraints derived from neutral meson mi%
		xing.
		Then, we show that left models can accomodate spontaneous CP violation when all quarks have Flavour Changing Neutr%
		al Couplings.
		Finally, we illustrate these concepts by considering a specific implementation with three Higgs doublets.
	\end{abstract}
\end{cover}

\section{Introduction}\lSc{beg}%%%%%%%%%%%%%%%%%%%%%%%%%%%%%%%%%%%%%%%%%%%%%%%%%%%%%%%%%%%%%%%%%%%%%%%%%%%%%%%%%%%%%%%%
Multi-Higgs Doublets Models (MHDMs) are common extensions of the Standard Model (SM) with multiple scalar doublets.
At first, they were introduced by T.~D.~Lee\cite{Lee:1973iz} to generate CP violation spontaneously when only two gene%
rations of matter were known.
Since then, MHDMs have become popular options for addressing baryogenesis\cite{Fromme:2006cm,Cline:1995dg,Cline:2011mm,
Dorsch:2016nrg} and dark matter\cite{LopezHonorez:2006gr,Goudelis:2013uca,LopezHonorez:2010eeh}.
Likewise, spontaneous CP Violation (CPV) now plays a critical role in the Nelson-Barr solution\cite{Nelson:1983zb,Barr%
:1984qx} for the strong CP problem.
However, MHDMs have tree-level Flavour Changing Neutral Couplings (FCNC) that are known to face several stringent expe%
rimental constraints\cite{Branco:2011iw}.
For that reason, Natural Flavour Conservation (NFC)\cite{Glashow:1976nt,Paschos:1976ay} is often introduced to remove %
all FCNC from the theory.
However, when all CPV is spontaneous, NFC gives rise to a real Cabibbo-Kobayashi-Maskawa (CKM) matrix\cite{Branco:1980%
sz} which has already been experimentally ruled out, including in the presence of new physics\cite{Botella:2005fc}.
Alternatively, Branco, Grimus and Lavoura (BGL)\cite{Branco:1996bq} introduced a class of Two-Higgs Doublets Models (2%
HDMs) with tree-level FCNC that are entirely fixed by the CKM matrix.
Detailed phenomenological analyses\cite{Botella:2015hoa,Botella:2014ska,Bhattacharyya:2014nja} have shown that they re%
main compatible with the current experimental data.
Unfortunately, when CPV is spontaneous, their CKM matrix is rendered real.
With that in mind, a generalization of BGL (gBGL) models has been proposed in\cite{Alves:2017xmk}.
A comprehensive phenomenological analysis of its quark sector\cite{Nebot:2018nqn} found that these models can remain in
agreement with the current experimental constraints when CPV is spontaneous as long as every quark possesses tree-level
FCNC.
In the leptonic extensions of gBGL models\cite{Alves:2021pak}, the complex phases in the Pontecorvo-Maki-Nakagawa-Saka%
ta (PMNS) and CKM matrices become related by a single vacuum phase.

In this paper, we seek to replicate the success of gBGL models by finding the largest class of MHDMs with the following
properties -- $(i)$ small number of flavour parameters; $(ii)$ naturally suppressed FCNC; $(iii)$ compatibility with s%
pontaneous CPV.
Due to the nature of these clauses, our analysis will be focused exclusively on the MHDM flavour sector.

In \rSc{mhdm}, we will settle the notation by reviewing the quark sector of the MHDM Yukawa Lagrangian.
Then, in \rSc{ufv} we will introduce the notion of Unitary Flavour Violation (UFV) before showing that MHDMs with it a%
re described by a small number of flavour parameters.
After that, in \rSc{lrm} we will proof that two of its subclasses, the left and right models, have naturally suppressed
tree-level FCNC, with the former being compatible with spontaneous CPV.
In \rSc{model}, we illustrate these concepts by considering a particular Three-Higgs Doublets Model (3HDM).
Finally, we present our conclusions in \rSc{end}.

\section{The General MHDM}\lSc{mhdm}%%%%%%%%%%%%%%%%%%%%%%%%%%%%%%%%%%%%%%%%%%%%%%%%%%%%%%%%%%%%%%%%%%%%%%%%%%%%%%%%%%%
The quark Yukawa sector of a generic MHDM is controlled by the following Lagrangian,\begin{math}\Eq{mhdm-Ls}{
		\Lg_Y=-\ab{q}^0_L\jf_aY_{da}d^0_R-\ab{q}^0_L\at{\jf}_aY_{ua}u^0_R+h.c.,
}\end{math}in which there are implicit sums over every scalar in the theory ($a=1,\dc,N$), $Y_{da},Y_{ua}$ describe $3%
\Mb3$ arbitrary complex matrices and $\at{\jf}_a=i\js_2\jf^*_a$.
After spontaneously breaking the electroweak symmetry with $\vev{\jf^T_a}=\fs{1}{\rs{2}}v_ae^{i\ja_a}\Sa(0\ms{9}1\Sa)$,
we make use of a real orthogonal matrix $O$ with $O_{1a}=v_a/v$ for $v^2=v^2_1+\db+v^2_N$ to define the Higgs basis be%
low,\begin{math}\Eq{mhdm-Bh}{
		H_a=O_{ab}\Sa(e^{-i\ja_b}\jf_b\Sa)\ev\mxb{C^+_a\\\fs{1}{\rs{2}}(v\jd_{a1}+R_a+iI_a)}\ms{-2}.
}\end{math}%
In it, the FCNC of a generic MHDM may be expressed in the following manner,\begin{math}\Eq{mhdm-Lh}{
		\Lg_Y&\sps-\fs{R_a}{v}\Sb[\ab{q}\Sa(N_{qa}\jg_R+N^\dg_{qa}\jg_L\Sa)q\Sb]-\fs{i\je_qI_a}{v}\Sb[\ab{q}\Sa(N_{qa}\jg_R
		-N^\dg_{qa}\jg_L\Sa)q\Sb],\\
%		&-\fs{\rs{2}}{v}\Sb\{C^+_a\Sb[\ab{u}\Sa(VN_{da}\jg_R-N^\dg_{ua}V\jg_L\Sa)d\Sb]+h.c.\Sb\}
}\end{math}where there are implicit sums over $q=d,u$ with $\je_d=-\je_u=1$, and we have introduced\begin{math}\Eq{mhd%
m-Ch}{
		N^0_{qa}=\fs{1}{\rs{2}}vO_{ab}e^{i\je_q\ja_b}Y_{qb},\qd N_{qa}=U^\dg_{qL}N^0_{qa}U_{qR}
}\end{math}after performing the Weak Basis (WB) transformations $q^0_X=U_{qX}q_X$ for $X=L,R$ that diagonalize the qua%
rk mass matrices $D^0_q\ev N^0_{q1}$.
As such, with the physical scalars\begin{math}\Eq{mhdm-Bp}{
		h_{\at{a}}=R_{\at{a}a}R_a+R_{\at{a},N+\ah{a}}I_{1+\ah{a}},
%		\qd h^+_{\ah{a}}=U_{\ah{a}\ah{b}}C^+_{1+\ah{b}}
}\end{math}where tilded (hatted) indices range from $1$ to $2N-1$ ($N-1$) with $R$ a real orthogonal matrix, the FCNC %
of a generic MHDM become controlled by the following Lagrangian,\begin{math}\Eq{mhdm-Lp}{
		\Lg_Y&\sps-\fs{h_{\at{a}}}{v}\Sb[\ab{q}\Sa(Z_{q\at{a}}\jg_R+Z^\dg_{q\at{a}}\jg_L\Sa)q\Sb],
%		-\fs{\rs{2}}{v}\Sb\{h^+_{\ah{a}}\Sb[\ab{u}\Sa(W_{d\ah{a}}\jg_R-W^\dg_{u\ah{a}}\jg_L\Sa)d\Sb]+h.c.\Sb\}\\
%		&-\fs{i\je_qG^0}{v}\Sa(\ab{q}D_q\jg_5q\Sa)-\fs{\rs{2}}{v}\Sb\{G^+\Sb[\ab{u}\Sa(VD_d\jg_R-D_uV\jg_L\Sa)d\Sb]+h.c.\Sb
%		\}
}\end{math}in which we have introduced the following couplings,\begin{math}\Eq{mhdm-Cp}{
		Z_{q\at{a}}&=R_{\at{a}a}N_{qa}+i\je_qR_{\at{a},N+\ah{a}}N_{q,1+\ah{a}}=R_{\at{a}1}D_q+\Sa(R_{\at{a},1+\ah{a}}+i\je_
		qR_{\at{a},N+\ah{a}}\Sa)N_{q,1+\ah{a}}.\\
%		W_{u\ah{a}}&=U_{\ah{a}\ah{b}}V^\dg N_{u,1+\ah{b}},\qd W_{d\ah{a}}=U^*_{\ah{a}\ah{b}}VN_{d,1+\ah{b}}
}\end{math}%
Since we have established that the flavour sector $\Lg_Y$ of a generic MHDM only contains $N-1$ non-trivial mass matri%
ces $N_{qa}$ and that $R$ is determined by the scalar potential, \rEq{mhdm-Cp} is described by $36(N-1)$ new flavour p%
arameters.

It is well-known that neutral meson systems provide the most stringent constraints of MHDMs.
As such, in this paper we will control their FCNC by demanding that the contributions of \rEq{mhdm-Lp} to the amplitude
of mixing in a $P^0=\ab{q}_iq_j$ system,\begin{math}\Eq{mhdm-Et}{
		M^P_{12}\Sa|_\mfa{TL}=\fs{5\Ju^2_Pf^2_Pm_P}{48m^2_hv^2}\an^{2N-1}_{\at{a}=1}n^{-2}_{\at{a}}\Sc[\Sa(Z^*_{q\at{a}}\Sa
		)^2_{ij}+\Sa(Z_{q\at{a}}\Sa)^2_{ji}-\fs{2}{5}\Sa(6-\Ju^{-2}_P\Sa)\Sa(Z^*_{q\at{a}}\Sa)_{ij}\Sa(Z_{q\at{a}}\Sa)_{ji}
		\Sc],
}\end{math}where $\Ju_P=m_P/(m_i+m_j)$ and $n_{\at{a}}=m_{\at{a}}/m_h$ with $h\ev h_1$ the $125\,\mfa{GeV}$ scalar det%
ected at the LHC, should remain below the current experimental observations, i.e.,\begin{math}\Eq{mhdm-Eg}{
		\Jd m_P\Sa|_\mfa{NP}=2\Sa|M^P_{12}\Sa|_\mfa{TL}<\Jd m_P.
}\end{math}%
As a result, in this paper MHDMs will be required to satisfy the following constraint,\begin{math}\Eq{mhdm-Ec}{
		\Sd|\an^{2N-1}_{\at{a}=1}n^{-2}_{\at{a}}\Sc[\Sa(Z^*_{q\at{a}}\Sa)^2_{ij}\ms{-2}+\ms{-1}\Sa(Z_{q\at{a}}\Sa)^2_{ji}%%
		\ms{-2}-\ms{-1}\fs{2}{5}\Sa(6\ms{-2}-\ms{-3}\Ju^{-2}_P\Sa)\Sa(Z^*_{q\at{a}}\Sa)_{ij}\Sa(Z_{q\at{a}}\Sa)_{ji}\Sc]\Sd
		|<\mfc{C}_P\ev\fs{24v^2m^2_h\Jd m_P}{5\Ju^2_Pf^2_Pm_P}.
}\end{math}%
In $K^0$, $D^0$, $B^0_d$ and $B^0_s$ systems, the right-hand side of this expression equates to\begin{math}\Eq{mhdm-En}
{
		\mfc{C}_K=\Sa(5.12\pm0.90\Sa)\ms{-2}\Mb10^{-5}\,\mfa{GeV}^2,&\qd\mfc{C}_{B_d}=\Sa(5.02\pm0.10\Sa)\ms{-2}\Mb10^{-3}%
		\,\mfa{GeV}^2,\\
		\mfc{C}_D=\Sa(1.66\pm0.20\Sa)\ms{-2}\Mb10^{-4}\,\mfa{GeV}^2,&\qd\mfc{C}_{B_s}=\Sa(1.189\pm0.022\Sa)\ms{-2}\Mb10^{-1
		}\,\mfa{GeV}^2.
}\end{math}\cp\nps%
Since the mass matrices $N_{qa}$ are linearly independent, in a natural MHDM where $|Y_{qa}|$ is order one \rEqs{mhdm-%
Ch}{mhdm-Cp} suggest that $|Z_{q\at{a}}|_{ij}\sm m_t$ should hold\ftn{.}
{Due to the linear independence, setting, for example, $(N_{d1})_{11}=m_d$ as no impact on $(N_{da\nq1})_{11}$.}
Thus, within those models \rEq{mhdm-Ec} states that some intricate combinations of squared order-one couplings are inf%
erior to $\mfc{C}_P/m^2_t$.
As such, in their $(sd,uc,bd,bs)$ sectors natural MHDMs require unnatural cancellations of about $(10^{-5},10^{-4},10^%
{-4},10^{-3})$ to avoid \rEq{mhdm-Ec}.

\section{Unitary Flavour Violation}\lSc{ufv}%%%%%%%%%%%%%%%%%%%%%%%%%%%%%%%%%%%%%%%%%%%%%%%%%%%%%%%%%%%%%%%%%%%%%%%%%%%
In this section, we will search for a large class of MHDMs with controlled FCNC.
These were previously shown to, as far as flavour is concerned, be fixed by the mass matrices $N_{qa}$.
As such, we are looking for MHDMs where $N_{qa}$ is related to SM parameters that, from dimensional analysis, must be %
in the quark mass matrices $D_q$.
Thus, we introduce UFV by claiming that MHDMs with it must contain a WB where the following relation is protected by an
Abelian symmetry,\begin{math}\Eq{ufv-Ds}{
		N^0_{qa}=L_{qa}D^0_qR_{qa},
}\end{math}in which $L_{qa},R_{qa}$ are matrices that do not contain flavour parameters.
Through \rEq{mhdm-Ch}, this expression may be rewritten into\begin{math}\Eq{ufv-Dy}{
		O_{ab}e^{i\je_q\ja_b}\Sa(Y_{qb}\Sa)_{ij}=O_{1b}e^{i\je_q\ja_b}\Sa(L_{qa}\Sa)_{im}\Sa(Y_{qb}\Sa)_{mn}\Sa(R_{qa}\Sa)_
		{nj},
}\end{math}where there are implicit sums over $b,m,n$.
Since all flavour parameters in this equation lie inside the Yukawa couplings and these are independent\ftn{,}
{All Abelian symmetries have a WB where they may be described by a field rephasing which either renders the Yukawa cou%
plings null or leaves them free.}
solving \rEq{ufv-Dy} for $L_{qa},R_{qa}$ is only possible when $Y_{qa}$ is dropped from it.
With that in mind, we must restrict UFV to remove the mixing of flavour indices by selecting diagonal $L_{qa}, R_{qa}$,
and thus write\begin{math}\Eq{ufv-Dr}{
		O_{ab}e^{i\je_q\ja_b}\Sa(Y_{qb}\Sa)_{ij}=l_{qa,i}r_{qa,j}O_{1b}e^{i\je_q\ja_b}\Sa(Y_{qb}\Sa)_{ij},
}\end{math}with $l_{qa,i}\ms{-2}\ev\ms{-2}(L_{qa})_{ii}$ and $r_{qa,j}\ms{-2}\ev\ms{-2}(R_{qa})_{jj}$.
Once more, we note that $Y_{qa}$ can only be dropped from this expression after the sum over $b$ is removed.
Clearly, that is only possible when\begin{math}\Eq{ufv-Dc}{
		\Sa(Y_{qa}\Sa)_{ij}\Sa(Y_{qb}\Sa)_{ij}\pp\jd_{ab}
}\end{math}holds.
When this condition applies, \rEq{ufv-Dr} gives rise to the following system,\begin{math}\Eq{ufv-Df}{
		l_{qa,i}r_{qa,j}=\fs{O_{ab}}{O_{1b}},
}\end{math}which has $N$ equations for each $(Y_{qb})_{ij}\nq0$.
Thankfully, this expression decouples into $2N$ independent systems with up to nine equations and five variables\ftn{.}
{For fixed $qa$, there are three variables in $l_{qa,i}$ and in $r_{qa,j}$.
However, only five impact their product.}
As such, while necessary \rEq{ufv-Dc} does not provide a sufficient condition for ensuring UFV.
Notice also that, since $O_{ab}$ is a function of vevs determined by the scalar potential, \rEq{ufv-Df} allows for $L_%
{qa},R_{qa}$ to be flavour-independent.
\cp
After performing the WB transformations which diagonalize $D_q$, \rEq{ufv-Ds} becomes\begin{math}\Eq{ufv-Ch}{\ms{-1}
		N_{qa}=\Sb(l_{qa,1}P^{U_{qL}}_1+l_{qa,2}P^{U_{qL}}_2+l_{qa,3}P^{U_{qL}}_3\Sb)D_q\Sb(r_{qa,1}P^{U_{qR}}_1+r_{qa,2}P^
		{U_{qR}}_2+r_{qa,3}P^{U_{qR}}_3\Sb),
}\end{math}where $P^X_i=X^\dg P_iX$ with $(P_k)_{ij}=\jd_{ik}\jd_{jk}$ are projection operators for $X$ unitary since %
\begin{math}\Eq{ufv-Po}{
		P^X_iP^X_j=P^X_i\jd_{ij},\qd P^X_1+P^X_2+P^X_3=1
}\end{math}applies.
After employing the CKM matrix $V=U^\dg_{uL}U_{dL}$ to establish the identity below,\begin{math}\Eq{ufv-Pv}{
		P^{U_{uL}}_iV=VP^{U_{dL}}_i,
}\end{math}we notice that just three projection operators contain every new parameter in \rEq{ufv-Ch}.
Since $P^{\Jl X}_i=P^X_i$ holds for $\Jl$ diagonal and unitary, each operator may be parametrized by the three real an%
gles and three complex phases in the unitary matrix\begin{math}\Eq{ufv-Pu}{
		U_{qX}=\ms{-3}\mxp{1\\&c_1&s_1\\&-s_1&c_1}\ms{-6}\mxp{c_2&&s_2e^{-i\ja}\\&1\\-s_2e^{i\ja}&&c_2}\ms{-6}\mxp{c_3&s_3%
		\\-s_3&c_3\\&&1}\ms{-6}\mxp{e^{i\jb_1}\\&e^{i\jb_2}\\&&1}\ms{-3}.
}\end{math}%
Thus, the flavour sector of a MHDM with UFV may only be described by 18 parameters.
Notice that such a number is independent on the amount of scalars $N$.
This was to be expected, however, since \rEq{ufv-Ds} implies that in MHDMs with UFV $D_q$ contains every flavour param%
eter.
Meanwhile, \rEq{ufv-Ch} suggests $(|Z_{d\at{a}}|_{ij},|Z_{u\at{a}}|_{ij})\sm(m_b,m_t)$.
Since this offers no natural suppression for the FCNC in their up sector, valid MHDMs with UFV are still fine-tunned to
the level of $10^{-4}$.

\section{Left and Right models}\lSc{lrm}%%%%%%%%%%%%%%%%%%%%%%%%%%%%%%%%%%%%%%%%%%%%%%%%%%%%%%%%%%%%%%%%%%%%%%%%%%%%%%%
In this section, we will consider two subclasses of UFV, the left and right models which were first introduced in\cite{
Alves:2018kjr}.
Since there is a strong analogy between the two, we shall focus on the former before highlighting its differences to t%
he latter.

Left models are, by definition, MHDMs with a WB in which the following relation where $L_{qa}$ is flavour-independent %
is protected by an Abelian symmetry,\begin{math}\Eq{lrm-Ds}{
		N^0_{qa}=L_{qa}D^0_q.
}\end{math}%
It follows that left models are MHDMs with UFV and $R_{qa}=1$.
Then, in them \rEq{ufv-Df} becomes the following system with $N$ equations for each $(Y_{qb})_{ij}\nq0$,\begin{math}\Eq
{lrm-Df}{
		l_{qa,i}=\fs{O_{ab}}{O_{1b}}.
}\end{math}%
Clearly, this expression will have no solutions unless the relation below holds,\begin{math}\Eq{lrm-Dc}{
		\Sa(Y_{qa}\Sa)_{ij_1}\Sa(Y_{qb}\Sa)_{ij_2}\pp\jd_{ab}.
}\end{math}%
When this applies, however, \rEq{lrm-Df} decouples into $2N$ independent systems, each with up to three distinct equat%
ions matched by three variables.
Thus, \rEq{lrm-Dc} is a necessary and sufficient condition for detecting left models which offers an alternative defin%
ition:
\begin{quote}
	``A left model is a MHDM with a WB where no row of the Yukawa couplings receives contributions from multiple scalar %
	doublets.''
\end{quote}\cp\nps
Likewise, right models can be defined as MHDMs with UFV and $L_{qa}=1$.
As a result, the argument above applies with $l_{qa,i}\ars r_{qa,j}$ and rows $\ars$ columns.

After diagonalizing the quark mass matrices, the FCNC of a left model are fixed by\begin{math}\Eq{lrm-Ch}{
		N_{qa}=\Sb(l_{qa,1}P^{U_{qL}}_1+l_{qa,2}P^{U_{qL}}_2+l_{qa,3}P^{U_{qL}}_3\Sb)D_q.
}\end{math}%
Thanks to \rEq{ufv-Pv}, we note that $N_{qa}$ only includes one independent projection operator.
As such, the flavour sector of any left model can be determined by six new parameters.
Meanwhile, in \rAx{fcnc} we show that all left models which satisfy\begin{math}\Eq{lrm-Ec}{
		\Sd(\an^{2N-1}_{\at{a}=1}\fs{1-R^2_{\at{a}1}}{m^2_{\at{a}}/m^2_h}\Sd)\Sd[\an^{N-1}_{\ah{a}=1}\Sb|l_{q,1+\ah{a},k}%%
		\Sb(P^{U_{qL}}_k\Sb)_{ij}\Sb|^2\Sd]<\mfc{C}'_P\ev\fs{24v^2m^2_h\Jd m_P}{5f^2_Pm^3_P}
}\end{math}avoid the experimental constraints of \rEq{mhdm-Ec}.
The left-hand side of this relation mixes parameters from the scalar potential ($l_{q,1+\ah{a},k},R_{\at{a}1},n_{\at{a}
} $) with flavour couplings ($U_{qL}$) and is naturally suppressed as \rEq{lrm-Ch} renders the FCNC proportional to the
largest valence quark mass in the neutral meson system.
The effect of the latter can be felt in\begin{math}\Eq{lrm-En}{
		\mfc{C}'_K=\Sa(5.324\pm0.027\Sa)\ms{-2}\Mb10^{-3},&\qd\mfc{C}'_{B_d}=\Sa(2.865\pm0.041\Sa)\ms{-2}\Mb10^{-4},\\
		\mfc{C}'_D=\Sa(1.03\pm0.12\Sa)\ms{-2}\Mb10^{-4},&\qd\mfc{C}'_{B_s}=\Sa(6.510\pm0.076\Sa)\ms{-2}\Mb10^{-3}.
}\end{math}%
Since the left-hand side of \rEq{lrm-Ec} envolves squared couplings, these results imply that left models only demand %
to be fine-tunned to the $(10^{-1},10^{-2},10^{-2},10^{-1})$ level in their $(sd,uc,bd,bs)$ sectors to avoid the exper%
imental constraints of \rEq{mhdm-Ec}.
Regarding right models, all arguments above apply with $l_{qa,i}\ms{-1}\ars r_{qa,j}$ and $U_{qL}\ars U_{qR}$, with the
exception of that used in the parameter counting.
Namely, their FCNC contain two independent projection operators, and thus may include 12 additional parameters.

When trivial CP is imposed as a good symmetry of the Lagrangian, there is a WB in which every Yukawa coupling is real.
Then,\begin{math}\Eq{lrm-Mg}{
		D^0_q=\Sa(e^{i\ja_{q_1}}P_1+e^{i\ja_{q_2}}P_2+e^{i\ja_{q_3}}P_3\Sa)\ah{D}^0_q
}\end{math}with $\ah{D}^0_q$ real while $q_i$ identifies the scalar which couples to the $i$-th row of the Yukawa coup%
lings follows directly from the alternate definition for left models.
After employing the polar decomposition of real matrices, $\ah{D}=O_1\Jl O^T_2$, where $O_1,O_2$ are real orthogonal m%
atrices while $\Jl$ is diagonal and real, \rEq{lrm-Mg} becomes\begin{math}\Eq{lrm-Md}{
		D^0_q=\Sa(e^{i\je_q\ja_{q_1}}P_1+e^{i\je_q\ja_{q_2}}P_2+e^{i\je_q\ja_{q_3}}P_3\Sa)O_{qL}D_qO^T_{qR}.
}\end{math}%
As such, the matrices which diagonalize a left model with spontaneous CPV look like\begin{math}\Eq{lrm-Uq}{
		U_{qL}=\Sa(e^{i\je_q\ja_{q_1}}P_1+e^{i\je_q\ja_{q_2}}P_2+e^{i\je_q\ja_{q_3}}P_3\Sa)O_{qL},\qd U_{qR}=O_{qR}.
}\end{math}%
At this stage, we note that the phases $\ja_{q_i}$ do not contribute to the projection operators $P^{U_{qL}}_i=O^T_{qL}
P_iO_{qL}$.
Thus, when CP is spontaneously broken the flavour sector of a left model can be parametrized by three real angles (che%
ck \rEq{ufv-Pu} with $\ja=\jb_1=\jb_2=0$).
Meanwhile, their mixing matrices $V$ and $V^R\ev U^\dg_{uR}U_{dR}$ are given by\begin{math}\Eq{lrm-Uv}{
		V=O^T_{uL}\Sb[e^{i(\ja_{u_1}+\ja_{d_1})}P_1+e^{i(\ja_{u_2}+\ja_{d_2})}P_2+e^{i(\ja_{u_3}+\ja_{d_3})}P_3\Sb]O_{dL},%
		\qd V^R=O^T_{uR}O_{dR}.
}\end{math}%
In\cite{Nebot:2018nqn}, it was shown that the texture of $V$ possesses a non-trivial complex phase when its middle com%
ponent is not proportional to the identity and all quarks have tree-level FCNC.
For right models, the discussion above is translated with $L\abs R$, such that the textures for their $V$ and $V^R$ be%
come swapped.
As such, it follows that only left models are compatible with spontaneous CPV.

\section{A left model with three doublets}\lSc{model}%%%%%%%%%%%%%%%%%%%%%%%%%%%%%%%%%%%%%%%%%%%%%%%%%%%%%%%%%%%%%%%%%%
In this section, we will study a 3HDM invariant under the following Abelian symmetry,\begin{math}\Eq{model-Ds}{
		\mxp{\jf_1\\\jf_2\\\jf_3}\ms{-3}\ars\ms{-3}\mxp{1\\&e^{i\jq}\\&&e^{-i\jq}}\ms{-6}\mxp{\jf_1\\\jf_2\\\jf_3}\ms{-3},%
		\qd q^0_L\ars\ms{-3}\mxp{1\\&1\\&&e^{i\jq}}\ms{-3}q^0_L,\qd d^0_R\ars d^0_R,\qd u^0_R\ars u^0_R,
}\end{math}with $\jq=2\jp/3$, and the trivial CP.
Here, the Yukawa couplings of this 3HDM look like\begin{math}\Eq{model-Dy}{
		Y_{d1}\sm Y_{u1}\sm\ms{-3}\mxp{\Mb&\Mb&\Mb\\\Mb&\Mb&\Mb\\0&0&0}\ms{-3},\qd Y_{d2}\sm Y_{u3}\sm\ms{-3}\mxp{0&0&0\\0&
		0&0\\\Mb&\Mb&\Mb}\ms{-3},\qd Y_{d3}=Y_{u2}=0,
}\end{math}where $\Mb$ represents an arbitrary real number.
Since these textures satisfy \rEq{lrm-Dc}, this 3HDM is a left model.
Then, in the Higgs basis below,\begin{math}\Eq{model-Dh}{
		O=\ms{-3}\mxp{v_1/v&v_2/v&v_3/v\\v_2/v'&-v_1/v'&0\\v_1/v''&v_2/v''&-v'^2/v_3v''}\ms{-3},
}\end{math}where $v'^2=v^2_1+v^2_2$ and $v''=vv'/v_3$, this left model is defined with \rEq{lrm-Df} through\begin{math}
\Eq{model-Dl}{
		\av{l}_{d1}=\Sa(1,1,1\Sa),\qd\av{l}_{d2}&=c^{-1}_2\Sa(t_1,t_1,-t^{-1}_1\Sa),\qd\av{l}_{d3}=t_2\Sa(1,1,1\Sa),\\
		\av{l}_{u1}=\Sa(1,1,1\Sa),\qd\av{l}_{u2}&=t_1c^{-1}_2\Sa(1,1,0\Sa),\qd\av{l}_{u3}=\Sa(t_2,t_2,-t^{-1}_2\Sa),
}\end{math}in which we introduced $v_1=vc_1c_2$, $v_2=vs_1c_2$ and $v_3=vs_2$ with $(c_i,s_i)\ev(\Cos\jq_i,\Sin\jq_i)$.
Notice that, being related to $v_{1,2,3}$, all $\av{l}_{qa}$ are fixed by the scalar potential.
Meanwhile, $\av{l}_{q1}$ is trivial since, by definition, $N_{q1}\ev D_q$, while the remaining $\av{l}_{qa}$ produce %%
\begin{math}\Eq{model-Ch}{
		N_{d2}=c^{-1}_2\Sb[t_1-\Sa(t_1+t^{-1}_1\Sa)P^{U_{dL}}_3\Sb]D_d&,\qd N_{d3}=t_2D_d,\\
		N_{u2}=t_1c^{-1}_2\Sb(1-P^{U_{uL}}_3\Sb)D_u&,\qd N_{u3}=\Sb[t_2-\Sa(t_2+t^{-1}_2\Sa)P^{U_{uL}}_3\Sb]D_u
}\end{math}thanks to \rEqs{ufv-Po}{lrm-Ch}.
Since $(P^{U_{dL}}_3)_{ij}=(U_{dL})^*_{3i}(U_{dL})_{3j}$, these mass matrices only see the third row of $U_{dL}$.
This happens because $l_{qa,1}=l_{qa,2}$ holds for all $q,a$.
Thus, after remembering that in left models with spontaneous CPV the projection operators $P^{U_{qL}}_i$ become real, %
we conclude that this flavour sector is fixed by just two real parameters,\begin{math}\Eq{model-Cp}{
		r_i\ev\Sa(O_{dL}\Sa)_{3i}=\Sa(c_{\ja_1}c_{\ja_2},c_{\ja_1}s_{\ja_2},s_{\ja_1}\Sa).
}\end{math}\cp\nps%
Meanwhile, after employing \rEq{ufv-Po} to simplify the following sums over $k=1,2,3$,\begin{math}\Eq{model-Es}{
		l_{d2,k}P^{U_{dL}}_k=c^{-1}_2\Sb[t_1-\Sa(t_1+t^{-1}_1\Sa)P^{U_{dL}}_3\Sb]&,\qd l_{d3,k}P^{U_{dL}}_k=t_2,\\
		l_{u2,k}P^{U_{uL}}_k=t_1c^{-1}_2\Sb(1-P^{U_{uL}}_3\Sb)&,\qd l_{u3,k}P^{U_{uL}}_k=t_2-\Sa(t_2+t^{-1}_2\Sa)P^{U_{uL}}
		_3,
}\end{math}we can make use of \rEq{lrm-Ec} to conclude that when the following bound is satisfied,\begin{math}\Eq{mode%
l-Ec}{
		r^2_ir^2_j\Mb\ms{-2}c^{-2}_2\Sa(t_1+t^{-1}_1\Sa)^2\Sd(\an^{2N-1}_{\at{a}=1}\fs{1-R^2_{\at{a}1}}{m^2_{\at{a}}/m^2_h}
		\Sd)<\mfc{C}'_{K,B_d,B_s},\\
		\ab{r}^2_1\ab{r}^2_2\Mb\ms{-2}\Sb[t^2_1c^{-2}_2+\Sa(t_2+t^{-1}_2\Sa)^2\Sb]\Sd(\an^{2N-1}_{\at{a}=1}\fs{1-R^2_{\at{a
		}1}}{m^2_{\at{a}}/m^2_h}\Sd)<\mfc{C}'_D,
}\end{math}with $\ab{r}_i\ev(O_{uL})_{3i}=V^*_{ij}r_j$, these left models avoid the experimental constraints related to
neutral meson mixing.
As shown in \rSc{lrm}, \rEq{model-Ec} only requires cancellations of a $10^{-2}$ degree.
In\cite{Alves:2020brq}, the scalar potential invariant under the $\mfn{Z}_3$ symmetry in \rEq{model-Ds} and the trivial
CP was shown to have a CP violating vacuum.
Thus, by using \rEq{lrm-Uv} to write\begin{math}\Eq{model-Ev}{
		V=e^{2i\ja_1}O^T_{uL}\Sb[P_1+P_2+e^{i(\ja_2+\ja_3-2\ja_1)}P_3\Sb]O_{dL}
}\end{math}we conclude that these left models can accomodate spontaneous CPV when all quarks have FCNC, i.e., if $r_{1,
2,3},\ab{r}_{1,2,3}\nq0$, and $\jq=2\jp/3$.

\section{Conclusions}\lSc{end}%%%%%%%%%%%%%%%%%%%%%%%%%%%%%%%%%%%%%%%%%%%%%%%%%%%%%%%%%%%%%%%%%%%%%%%%%%%%%%%%%%%%%%%%%
In \rSc{mhdm}, we recalled the FCNC of a generic MHDM to show that they are described by too many parameters while also
facing stringent experimental constraints.
In order to get an handle on the latter, we adopted a simplified approach by looking exclusively at neutral meson mixi%
ng.
In \rSc{ufv}, we began to address these issues with UFV.
By forcing all flavour couplings to lie within the quark mass matrices, UFV allows for just three unitary matrices to %
contain every new parameter in the Yukawa Lagrangian.
As such, we found that the flavour sector of a MHDM with UFV can always be described by up to 18 parameters, thus solv%
ing the first issue.
Unfortunately, the little suppression offered by UFV to the FCNC (a factor of $y_b=m_b/v$ in the down sector) proofed %
to be insufficient.
For that reason, in \rSc{lrm} we studied two of its subclasses, the left and right models.
By suppressing the FCNC for neutral meson systems with their heaviest valence quark mass, both rendered the fine-tunni%
ng required to avoid the experimental constraints acceptable.
Left and right models were, however, found to be distinguished by two of its properties.
Firstly, the flavour sector of the former may only be described by six new parameters while that of the latter can 
be the owner of twelve.
Secondly, we found that only left models may accomodate spontaneous CPV.
As such, we illustrated these concepts in \rSc{model} by considering a left model with three Higgs doublets.
After verifying that we were studying a left model by constructing its Yukawa textures with an Abelian symmetry, we fo%
und that its FCNC were fixed by two real parameters that only need to be fine-tunned to the $10^{-2}$ level in order to
avoid the constraints employed.
Then, we showed that, when the symmetry was made to be a $\mfn{Z}_3$, their scalar potential can generate spontaneous %
CPV which we transported into a complex CKM.

\section*{Acknowledgments}%%%%%%%%%%%%%%%%%%%%%%%%%%%%%%%%%%%%%%%%%%%%%%%%%%%%%%%%%%%%%%%%%%%%%%%%%%%%%%%%%%%%%%%%%%%%%
We would like to thank M. Nebot and F. J. Botella for some helpful discussions.
Also, the author has received support from \tfi{Generalitat Valenciana} under the grant CIPROM 2022-36.
\dBib{ufv}

\section{Experimental Data}\lSc{exp}%%%%%%%%%%%%%%%%%%%%%%%%%%%%%%%%%%%%%%%%%%%%%%%%%%%%%%%%%%%%%%%%%%%%%%%%%%%%%%%%%%%
Throughout the previous sections, we have employed the experimental results displayed in the tables below
together with $G_F=(1.1663788\pm0.0000006)\ms{-1}\Mb10^{-5}\,\mfa{GeV}^{-2}$\cite{ParticleDataGroup:2022pth}
and $m_h=(125.25\pm0.17)\,\mfa{GeV}$.
\begin{tab}{exp-PDG}{cc|cc}{Quark masses taken from\cite{ParticleDataGroup:2022pth}}{
		Observable&Value&Observable&Value\\\hline
		$m_u$ ($\mfa{MeV}$)&$2.16\pm0.49$&$m_d$ ($\mfa{MeV}$)&$4.67\pm0.48$\\
		$m_c$ ($\mfa{GeV}$)&$1.27\pm0.02$&$m_s$ ($\mfa{MeV}$)&$93.4\pm8.6$\\
		$m_t$ ($\mfa{GeV}$)&$172.69\pm0.30$&$m_b$ ($\mfa{GeV}$)&$4.18\pm0.03$\\
}\end{tab}
\begin{tab}{exp-NM}{c|ccc}{Observables related to neutral meson systems -- $m_P$ and $\Jd m_P$ experimentally measured%
\cite{ParticleDataGroup:2022pth}, $f_P$ determined from lattice calculations with $N_f=2+1+1$\cite{FlavourLatticeAvera%
gingGroupFLAG:2021npn}}{
		P&$m_P$ ($\mfa{MeV}$)&$\Jd m_P$ ($\mfa{MeV}$)&$f_P$ ($MeV$)\\\hline
		$K$&$497.611\pm0.013$&$(3.484\pm0.006)\ms{-2}\Mb10^{-12}$&$155.7\pm0.3$\\
		$D$&$1864.84\pm0.05$&$(6.56\pm0.76)\ms{-2}\Mb10^{-12}$&$212.0\pm0.7$\\
		$B_d$&$5279.66\pm0.12$&$(3.334\pm0.013)\ms{-2}\Mb10^{-10}$&$190.0\pm1.3$\\
		$B_s$&$5366.92\pm0.10$&$(1.1693\pm0.0004)\ms{-2}\Mb10^{-8}$&$230.3\pm1.3$\\
}\end{tab}

\section{Managing the FCNC of a left model}\lSc{fcnc}%%%%%%%%%%%%%%%%%%%%%%%%%%%%%%%%%%%%%%%%%%%%%%%%%%%%%%%%%%%%%%%%%%
In this appendix, we will process the FCNC of a left model.
Thus, we start by combining \rEqs{mhdm-Cp}{lrm-Ch} to write\begin{math}\Eq{fcnc-Cp}{
		\Sa(Z_{q\at{a}}\Sa)_{ij}=m_j\Sa(R_{\at{a},1+\ah{a}}+i\je_qR_{\at{a},N+\ah{a}}\Sa)l_{q,1+\ah{a},k}\Sb(P^{U_{qL}}_k%%
		\Sb)_{ij},
}\end{math}with implicit sums over $\ah{a}=1,\dc,N-1$ and $k=1,2,3$.
In the neutral meson systems considered, one valence quark mass dominates.
As such, after replacing \rEq{fcnc-Cp} inside \rEq{mhdm-Ec} we find that left models should satisfy\ftn{~}
{In \rEq{fcnc-Eg}, we considered $m_j\gg m_i$.
This choice has no impact on our argument.}
\begin{math}\Eq{fcnc-Eg}{
		\Sd|\an^{2N-1}_{\at{a}=1}n^{-2}_{\at{a}}\Sa(Z^*_{q\at{a}}\Sa)^2_{ij}\Sd|<\fs{24v^2m^2_h\Jd m_P}{5\Ju^2_Pf^2_Pm_P}.
}\end{math}%
At this stage, we recall the triangle inequality for complex numbers,\begin{math}\Eq{fcnc-Mt}{
		\Sb|\An_nz_n\Sb|\le\an_n\Sa|z_n\Sa|,
}\end{math}to produce\begin{math}\Eq{fcnc-Et}{
		\Sd|\an^{2N-1}_{\at{a}=1}n^{-2}_{\at{a}}\Sa(Z^*_{q\at{a}}\Sa)^2_{ij}\Sd|\le\an^{2N-1}_{\at{a}=1}n^{-2}_{\at{a}}\Sa|
		Z_{q\at{a}}\Sa|^2_{ij}.
}\end{math}%
Then, we make use of the famous Cauchy-Schwarz inequality for vector products,\begin{math}\Eq{fcnc-Mv}{
		\Sb|\An_nu_nv^*_n\Sb|^2\le\Sb(\An_n|u_n|^2\Sb)\Sb(\An_n|v_n|^2\Sb),
}\end{math}while identifying $u_{\ah{a}}=l_{q,1+\ah{a},k}(P^{U_{qL}}_k)_{ij}$ and $v_{\ah{a}}=R_{\at{a},1+\ah{a}}-i\je_
qR_{\at{a},N+\ah{a}}$ to derive\begin{math}\Eq{fcnc-Ev}{
		\Sa|Z_{q\at{a}}\Sa|^2_{ij}&\le m^2_j\Sd(\an^{N-1}_{\ah{a}=1}R^2_{\at{a},1+\ah{a}}+R^2_{\at{a},N+\ah{a}}\Sd)\an^{N-1
		}_{\ah{a}=1}\Sb|l_{q,1+\ah{a},k}\Sb(P^{U_{qL}}_k\Sb)_{ij}\Sb|^2\\
		&=m^2_j\Sa(1-R^2_{\at{a}1}\Sa)\an^{N-1}_{\ah{a}=1}\Sb|l_{q,1+\ah{a},k}\Sb(P^{U_{qL}}_k\Sb)_{ij}\Sb|^2.
}\end{math}%
After combining \rEqs{fcnc-Et}{fcnc-Ev} to obtain the expression below,\begin{math}\Eq{fcnc-Mc}{
		\Sd|\an^{2N-1}_{\at{a}=1}n^{-2}_{\at{a}}\Sa(Z^*_{q\at{a}}\Sa)^2_{ij}\Sd|\le m^2_j\Sd(\an^{2N-1}_{\at{a}=1}\fs{1-R^2
		_{\at{a}1}}{n^2_{\at{a}}}\Sd)\Sd[\an^{N-1}_{\ah{a}=1}\Sb|l_{q,1+\ah{a},k}\Sb(P^{U_{qL}}_k\Sb)_{ij}\Sb|^2\Sd],
}\end{math}we conclude that every left model which satisfies the following relation respects \rEq{mhdm-Ec},\begin{math}
\Eq{fcnc-Ec}{
		\Sd(\an^{2N-1}_{\at{a}=1}\fs{1-R^2_{\at{a}1}}{n^2_{\at{a}}}\Sd)\Sd[\an^{N-1}_{\ah{a}=1}\Sb|l_{q,1+\ah{a},k}\Sb(P^{U
		_{qL}}_k\Sb)_{ij}\Sb|^2\Sd]<\fs{24v^2m^2_h\Jd m_P}{5f^2_Pm^3_P}.
}\end{math}%

\end{document}